\documentclass[12pt,aps,prb]{revtex4}

\usepackage{amsmath}    
\usepackage{graphicx}   

\newcommand{\NiFe}{Ni$_{81}$Fe$_{19}$}
\newcommand{\CoFe}{Co$_{90}$Fe$_{10}$}
\newcommand{\dR}{\ensuremath{\Delta{R}}}
\newcommand{\SiN}{SiN}
\newcommand{\ds}{$\delta_{s}$}
\newcommand{\Ohm}{$\Omega$}
\newcommand{\Rb}{$R_B$}
\newcommand{\Rn}{$R_{NL}$}

\newcommand{\rhoN}{$\rho_{N}$}

\begin{document}

\title{Film Edge Nonlocal Spin Valves}
\author{Andrew T. McCallum and Mark Johnson}
 \affiliation{Naval Research Laboratory, Washington, DC 20375, USA}
 \email{mccallum@anvil.nrl.navy.mil}   

\begin{abstract}

Spintronics is a new paradigm for integrated digital electronics.
Recently established as a niche for nonvolatile magnetic random
access memory (MRAM), it offers new functionality while
demonstrating low power and high speed performance. However, to
reach high density spintronic technology must make a transition to
the nanometer scale. Prototype devices are presently made using a
planar geometry and have an area determined by the lithographic
feature size, currently about 100 nm. Here we present a new
nonplanar geometry in which one lateral dimension is given by a film
thickness, the order of 10 nm. With this new approach, cell sizes
can  shrink by an order of magnitude. The geometry is demonstrated
with a nonlocal spin valve, where we study devices with an
injector/detector separation much less than the spin diffusion
length.

\end{abstract}

\maketitle

Spintronics is a unique approach to information processing that
utilizes the spin state of a conduction electron, rather than its
charge\cite{zutic}. The dominant device families include two
sandwich structures, the giant magnetoresistance (GMR) spin valve
and the magnetic tunnel junction (MTJ),  and the lateral spin valve
which is described below. All devices have operational similarity.
Their structure includes two single domain ferromagnetic layers, F1
and F2, that are separated by a nonmagnetic layer, N. The
magnetization M1 of F1 is typically  ``pinned,'' constrained to lie
in a given direction. Layer F2, called the free layer, is created in
a way that its magnetization, M2, has a uniaxial anisotropy such
that M2 can be oriented parallel or antiparallel with M1. GMR spin
valves and MTJs are characterized by a low (high) electric
resistance when M1 and M2 are parallel (antiparallel). The
difference between resistive values of the parallel and antiparallel
configurations, relative to the former, is the magnetoresistance
ratio (MR) that characterizes the device. The N layer of an MTJ is a
thin dielectric tunnel barrier, and that of a GMR spin valve is a
thin layer of nonmagnetic metal such as copper. For a lateral spin
valve, the N layer  is a thin nonmagnetic metal wire. However, the
electric resistance is low (high) when M1 and M2 are antiparallel
(parallel).

All these different devices are fabricated using standard top down
lithographic processing. Prototype spin valves and MTJs are
fabricated as thin film layered stacks with elliptical shape and
dimensions of roughly 70 nm by 130 nm, and have shown excellent
characteristics. However, problems with magnetic anisotropies that
may arise when dimensions shrink below these values are one reason
why prototypes with smaller areal dimensions have found less success
for digital electronics applications. Recently, a spin valve with
diameter of 40 nm had a successful demonstration as a microwave
resonator \cite{Ansermet}.

Nonlocal lateral spin valves (NLSVs) were first fabricated in
1985\cite{JohnsonSilsbee1985}. Recently, numerous groups have
created devices using many different kinds of metals and
semiconductors\cite{haviland,Holub2006,Lou2007,valenzuela2004,Tombros2007}.
Lateral spin valves have been fabricated with dimensions of order
100 nm and have been used to resolve basic issues of spin transport,
such as the spatial separation of spin and charge currents in a
nonlocal geometry\cite{haviland}. They can be used to study
fundamental properties such as electron spin flip scattering rates
and spin polarization properties of ferromagnet / nonmagnet (F/N)
interfaces, properties that are relevant to all spintronic
devices\cite{Stiles1996,kelly2006,Butler1997}. Furthermore, the fact
that the resistance changes in NLSVs increase with decreasing device
size may make these devices practical for future MRAM or sensor
applications.

\begin{figure}[h]
  \begin{center}
    \includegraphics[width=85mm,angle=0]{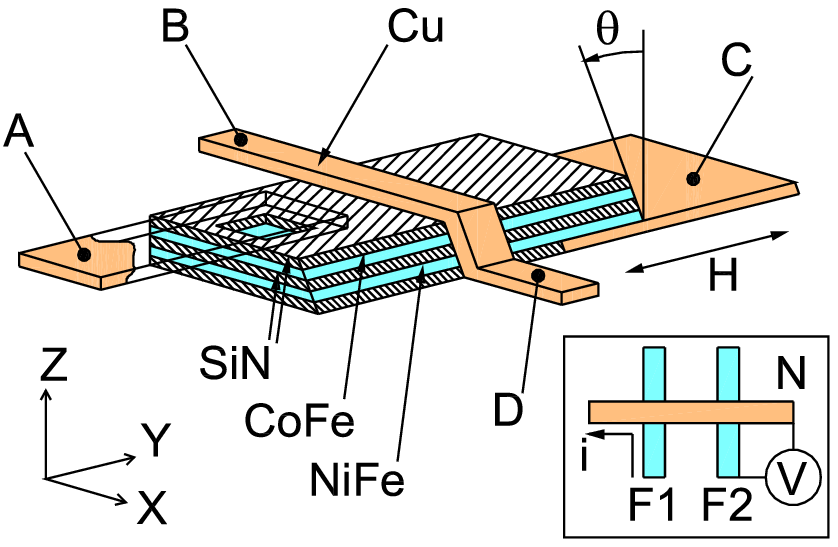}
    \caption[Device MockUp]{
    (Color online) A perspective diagram of a FENL spin valve. The distance between
    the ferromagnetic contacts along the Cu channel is defined by the thickness of the
    insulator (SiN) separating the two layers.  Nonlocal resistance
    measurements are made either by injecting current at
    contact A, with a ground at B, and measuring the voltage between contacts C and D or injecting current
    at contact C, with a ground at D, and measuring the voltage between contacts A and
    B. Inset: Top view of a lateral spin valve.}
    \label{fig:DeviceDiagram}
  \end{center}
\end{figure}

Previous NLSVs have had the same planar geometry, shown in the inset
of Fig. \ref{fig:DeviceDiagram}. Two or more ferromagnetic (FM)
electrodes are deposited on top of or under a narrow nonmagnetic
channel (N). These devices are often referred to as lateral spin
valves because the FM layers are arranged across the surface of the
substrate and not on top of one another as in a  GMR structure. When
a current $I$ is driven from one of the ferromagnetic electrodes,
the injector (F1), a spin accumulation forms in the channel near F1.
If the other ferromagnetic electrode, the detector (F2), is close
enough to F1 a voltage difference will form across the F2/N
interface in response to the spin accumulation. While this voltage
difference can not be measured directly, a voltage $V$ can be
measured between F2 and a region of the channel far from the current
path, and $V$ includes the voltage difference across the F2/N
interface. To isolate the effect of the spin accumulation we measure
the nonlocal resistance \Rn$\equiv V/I$ with the magnetizations M1
and M2 parallel and antiparallel to one another. The nonlocal
resistance change \dR$\equiv$\Rn{(Parallel)}-\Rn{(Antiparallel)} is
due to the effect of spin accumulation and is insensitive to the
Hall and anisotropic magnetoresistance effects.

The nonlocal resistance change of many NLSVs is proportional to
$e^{-L/\delta_s}$, where $L$ is the center to center distance
between F1 and F2 and \ds\ is the spin diffusion
length\cite{JohnsonSilsbee1985,Johnson1994,valenzuela2004}. This
exponential dependence is caused by spin flip scattering of
electrons in the channel. For a lateral spin valve $L$ must be
greater than the minimum (half pitch) lithographic feature size $f$.
Since $f$ is the same order of magnitude as \ds\ in many materials,
$e^{-L/\delta_s}$ is much less than 1 and \dR\ is significantly
limited.

In this Letter we introduce a nonplanar geometry for NLSVs where the
injector/detector spacing $L$ is much less than $\delta_s$ and much
smaller than in lateral spin valves. By fabricating the device
across the edge of a multilayered film stack, the critical dimension
(the spacing $L$) is determined by the thickness of one of the
layers in the stack, the order of 10 nm. The area of our NLSV
device, given approximately by the product of $L$ and wire width
$w$, is an order of magnitude smaller than that of state of the art
planar NLSVs\cite{haviland,valenzuela2004}.

These film edge nonlocal (FENL) spin valve devices were fabricated
using the edge of a lithographically patterned multilayer film,
formed by ion milling with a photoresist mask (Fig.
\ref{fig:DeviceDiagram}). Creating the edge by a removal process
allowed the individual layers of the multilayer to terminate cleanly
against the edge. The multilayer film consisted of two FM layers
that act as F1 and F2, separated and capped with insulating layers.
Electrical contact was made separately to each FM layer. Then a
narrow metal wire, that acts as the channel, was deposited across
the edge of the multilayer film. An examination of the geometry
reveals that the center to center spacing between the FM layers can
be expressed as $L=(\bar{t}_{FM} + t_{I})/cos(\theta)$, where
$\bar{t}_{FM}$ is the average thickness of the two FM layers,
$t_{I}$ is the thickness of the insulating layer between the FM
layers and $\theta$ is the angle the edge makes from the z-axis.
While the geometry of these FENL spin valves is very different from
planar lateral spin valves, the fundamental physics describing the
flow of electrons through the devices is the same. These FENL spin
valves do not include any tunnel barriers and are different from
\textquotedblleft{edge} junctions\textquotedblright\ previously
described\cite{Heiblum1978}.

\begin{figure}[h]
  \begin{center}
    \leavevmode
    \includegraphics[width=85mm]{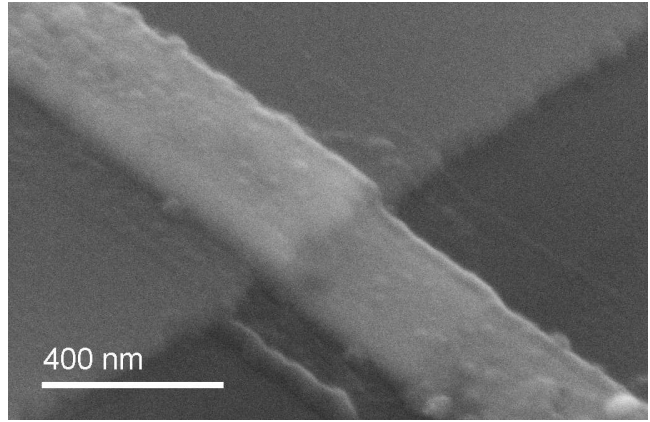}
    \caption[SEM]{
        An SEM image of a FENL spin valve taken at a 45 degree angle from the direction perpendicular to the sample
        plane. Note that the small thickness of the edge provides two independent and isolated electrical contacts to separate injector and detector films.}
    \label{fig:SEMofDevice}
  \end{center}
\end{figure}

The materials used in the devices and their thicknesses were chosen
for their electrical and magnetic properties. Starting at the
bottom, the multilayer consisted of 10 nm of \NiFe, 27 nm of \SiN,
10 nm of \CoFe, and 27 nm of \SiN\ on top (refer to Fig.
\ref{fig:DeviceDiagram}). The FM and \SiN\ layers were sputter
deposited at a pressure of 2.5 mTorr and 15 mTorr respectively. The
FM layers were thick enough to have low coercivities yet thin enough
to minimize $L$. The \SiN\ layers were thick enough so that there
were no pinholes between the NiFe and CoFe layers. This was verified
by measuring the resistance between these two layers before the Cu
wire was deposited. In our devices M1 is not pinned. However, the
NiFe and CoFe layers have different coercivities so that  sweeping a
magnetic field in the y-direction can change the magnetizations of
the two layers between parallel and antiparallel orientations. A
uniaxial anisotropy in both ferromagnetic layers was induced by
depositing these layers in a magnetic field along the y-axis. The
uniaxial anisotropy increased the squareness of the hysteresis loops
and allowed the magnetizations to be antiparallel over a wide field
range. Based on scanning electron microscope (SEM) images of the
devices, like the image shown in Fig. \ref{fig:SEMofDevice}, it was
determined that $\theta$=$(30\pm5)^{\circ}$ giving $L$=$(42\pm2)$
nm. The Cu wires used as channels in these devices were deposited by
electron beam evaporation and pattered by electron beam lithography
to be 330 nm wide and 50 nm thick. The edges of the ferromagnetic
layers were cleaned by ion milling immediately before depositing the
Cu wires to remove any oxide. The resistance area product of the
interface between the Cu channel and each of the FM layers is
approximately $5 \times 10^{-3}$ $\Omega \mu$m$^2$, much less than
for a tunnel junction.

\begin{figure}[h]
  \begin{center}
    \leavevmode
    \includegraphics[width=85mm]{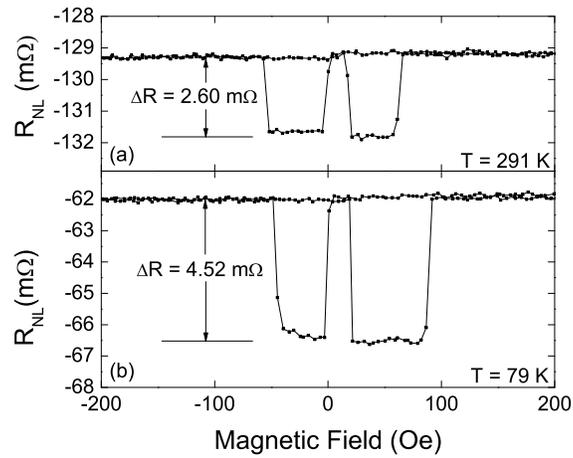}
    \caption[Hysteresis loop for device with Cu channel]{
        Nonlocal resistance data  from FENL spin valve device 15 on chip \textquotedblleft{B}\textquotedblright\ as a
        function of applied magnetic field at 291 K (a) and 79 K (b).}
    \label{fig:CuHysteresisLoop}
  \end{center}
\end{figure}

Transport measurements were made with a lock-in amplifier using bias
currents of approximately 0.5 mA at a frequency of 10 Hz. The
magnetic fields, used to change the magnetizations of the two
ferromagnetic layers, were applied along the y-axis. Nonlocal
resistance data taken from a FENL spin valve device are shown in
Fig. \ref{fig:CuHysteresisLoop}. Here $I$ is driven from A to B and
$V$ is measured from C to D (refer to Fig. \ref{fig:DeviceDiagram})
so that the CoFe acts as F1 and the NiFe acts as F2. The data show
clear differences between \Rn\ in the parallel and antiparallel
magnetization states making \dR\ easily measurable. The magnetic
properties of seven nominally identical devices were quite similar,
and the magnitude of \dR\ at 291 K varied from 1.4 m\Ohm\ to 2.60
m\Ohm.

For quantitative analysis, a comparison must first be made between
the interface resistance $R_i$ and the characteristic resistances
$R_N \equiv \rho_N \delta_s / A_N $ and $R_F \equiv \rho_F \delta_F
/ A_F$, where $\rho_N$, $\rho_F$ and $\delta_s$, $\delta_F$ are the
resistivities and spin diffusion lengths of the nonmagnetic and
ferromagnetic (F) metals, $A_N$ is the cross-sectional area of the
$N$ channel, and $A_F$ is the area of the interface between injector
(or detector) and the channel\cite{Johnson1987}. In our devices, we
measure $1.7 \ \Omega \approx  R_i >> R_N = 0.8 \ \Omega$ and $R_F /
R_N \approx 1$. These devices are not in a regime of negligible
interface resistance and we use Johnson-Silsbee theory to relate
\dR\ to empirically measurable parameters through the expression:
\begin{equation}\label{DeltaR}
    \Delta{R} = \frac{\eta^2\rho_N \ \delta_s}{A_N}e^{-L/\delta_s},
\end{equation}
where $\eta$ is the geometric mean of the polarizations of the two
ferromagnetic interfaces\cite{Johnson1994,valenzuela2004}.

The geometry of FENL spin valves prevents the fabrication of devices
with $L>\delta_s$, precluding the possibility of determining \ds\
from fits of the form $\Delta{R}(L)$. A comparison to theory can be
made using a value for \ds\ found by other researchers studying Cu
with similar resistivity. At room temperature, published values of
\ds\ are 400 nm in Cu with $\rho_N$=2.3
$\mu\Omega$cm\cite{Kimura2008} and 500 nm in Cu with $\rho_N$=2
$\mu\Omega$cm\cite{Kimura2005} and 5.5
$\mu\Omega$cm\cite{Garzon2005}. The resistivity of the Cu in our
devices was $\rho_N$=3.0$\pm$0.2 $\mu\Omega$cm, similar to Cu used
by these other researchers. Assuming a median value,
\ds=(450$\pm$50) nm, we find a range of values of $\eta$ from
$\eta$=(5.9$\pm$0.4)\% for the interfaces in the device with
\dR=(2.60$\pm$0.05) m\Ohm\ to $\eta$=(4.3$\pm$0.4)\% for the
interfaces in the device with \dR=(1.4$\pm$0.1) m\Ohm. These values
of $\eta$ lay in the middle of the range of $\eta$, from
2\%\cite{Hoffmann2007} up to 12\%\cite{godfrey2006}, published for
other all-metal NLSVs with low (but non-negligible) resistance ohmic
interfaces, at room temperature.

To further investigate the properties of our FENL spin valves we
measured three devices at cryogenic temperatures [see Fig.
\ref{fig:CuHysteresisLoop}(b)]. The ratio of \dR\ at room
temperature to that measured at 79 K varied between 52\% and 58\%
for these three devices even though the room temperature value was
different for each device. The ratio measured in our devices was
significantly higher than ratios measured in other published studies
of planar NLSVs with Cu channels, where the ratio was 38\% for
devices with $L$=250 nm and 15\% for devices with $L$=1650
nm\cite{Kimura2008}. In our FENL spin valves, $L$=42 nm is much
shorter than \ds\ at both room and cryogenic temperatures.
Therefore, we have $e^{-L/\delta_s}\sim{1}$ at both temperatures and
the exponential term contributes little to the temperature
dependence of \dR.

Another way way to characterize our FENL spin valves is to measure
the baseline resistance
\Rb$\equiv$(\Rn{(Parallel)}+\Rn{(Antiparallel)})/2. The baseline
resistance depends on $L$, \rhoN, the width of the channel and the
homogeneity of the interface resistances\cite{Johnson2007}. For our
devices,  theory\cite{Johnson2007} predicts a range of possible
values for \Rb,  between 0 \Ohm\ if the interface resistances are
homogeneous and -0.1 \Ohm\ if the interface resistances are
characterized by point contacts on opposite sides of the channel.
Measured values of \Rb\ for all seven devices were negative, as
predicted by theory, but only one had a measured \Rb\ in the
predicted range. The remaining devices had values between -0.13
\Ohm\ and -7.35 \Ohm. The discrepancy between calculation and
experiment is not understood, but may involve factors related to the
unusual nonplanar geometry or the roughness of the edge (refer to
Fig. 2). We note that while \Rb\ varied by two orders of magnitude
in these seven devices the values of \dR\ were all in the narrow
range 2.0 $\pm$ 0.6 m\Ohm.

There are several possible uses for FENL spin valves which take
advantage of their unique properties. One promising possibility is
to investigate spin momentum transfer effects\cite{Ansermet}. The
torque delivered by spin momentum transfer will act on a small
volume of the FM material near the edge in these devices, possibly
lowering the critical currents for switching and making them
attractive for MRAM applications. Another use for these devices is
to study the magnetization dynamics at the edge of ferromagnetic
films, which are important to understand switching in small
elements\cite{maranville2006,Jorzick2002}. FENL spin valves are
uniquely suited to observe these edge modes because the nonlocal
resistance of the devices is determined by the direction of the
magnetization within a few nanometers of the edge.

More generally, our data clearly demonstrate that spin polarized
current can be injected efficiently from the edge of a thin
ferromagnetic film. The film thickness of our prototypes, $d_F = 10$
nm, could presumably be made much smaller,  $d_F \sim 1$ nm. This
suggests interesting possibilities. A ferromagnetic layer could be
fabricated in a trench, in a plane perpendicular to the surface of a
chip. The film edge could be incorporated as part of a current
perpendicular to the plane (CPP) spin valves. Such a film edge CPP
spin valve would be expected to have the relatively large GMR of
planar CPP spin valves with a lower resistance than MTJs. The cell
area could be an order of magnitude smaller than that of either of
these planar devices.

In conclusion we have introduced a new nonplanar geometry that uses
the edge of a multilayer film, and adapted it to fabricate lateral
nonlocal spin valves in which the distance between ferromagnetic
electrodes is the order of ten nanometers. This enabled a study of
spin accumulation in the unexplored size regime where the electrode
spacing $L$ is much smaller than the spin diffusion length. The
nonlocal resistance changes in these devices are consistent with
Johnson-Silsbee theory. These results demonstrate that spin can be
injected from the edge a film through narrow interfaces about 30
atoms wide. The relatively large resistance changes, small interface
areas and the edge specific nature of their operation may make FENL
spin valves useful for fundamental studies and applications.

{\bf Acknowledgement}: The authors are grateful for the support of
the Office of Naval Research, funding document N0001408WX20705, and
they gratefully acknowledge use of the facilities at the NRL
Nanoscience Institute.

\newpage
\renewcommand{\baselinestretch}{2}

\newpage

\renewcommand{\baselinestretch}{1.5}


\end{document}